%% file: main.tex
\documentclass[submission,copyright,creativecommons]{eptcs}
\providecommand{\event}{SYNT 2017} % Name of the event you are submitting to
\usepackage{breakurl}             % Not needed if you use pdflatex only.
\usepackage{underscore}           % Only needed if you use pdflatex.
\usepackage{multirow}
\usepackage{xspace} 
\usepackage{hyperref} 
\usepackage{color} 
\usepackage{amsmath,amsthm}

\input{macros}

\newtheorem{dfn}{Definition}
\newtheorem{example}{Example}

\title{SyGuS Techniques in the Core of an SMT Solver}
\author{Andrew Reynolds
\institute{University of Iowa\\
Iowa City, Iowa, USA}
\email{andrew.j.reynolds@gmail.com}
\and
Cesare Tinelli
\institute{University of Iowa\\
Iowa City, Iowa, USA}
\email{tinelli@uiowa.edu}
}
\def\titlerunning{SyGuS Techniques in the Core of an SMT Solver}
\def\authorrunning{A. Reynolds \& C. Tinelli}
\begin{document}
\maketitle

\begin{abstract}
We give an overview of recent techniques for implementing
syntax-guided synthesis (SyGuS) algorithms
in the core of Satisfiability Modulo Theories (SMT) solvers.
We define several classes of synthesis 
conjectures and corresponding
techniques that can be used when dealing
with each class of conjecture.
\end{abstract}

\section{Introduction}

A synthesis conjecture asks whether there exists a structure
for which some property holds universally.
Traditionally, such conjectures are very challenging 
for automated reasoners.
Syntax-guided synthesis (or SyGuS) is a recently
introduced paradigm~\cite{AlurETAL13SyntaxguidedSynthesis} 
where a user may provide syntactic hints to
guide an automated reasoner in its search for solutions to synthesis conjectures.
A number of recent solvers based on this paradigm have been successfully used
in applications, including
correct-by-construction program snippets~\cite{SolarLezama13ProgramSketching}
and for implementation of distributed protocols~\cite{udupa2013transit}.

Satisfiability Modulo Theories (SMT) solvers
have historically been used as subroutines 
for automated synthesis tasks~\cite{gulwani2011synthesis,tiwari2015program,alurscaling,DBLP:conf/aips/WangDCK16}.
More recently, we have advocated in Reynolds et al.~\cite{ReynoldsDKBT15Cav} 
for SMT solvers to play a more active role in solving synthesis conjectures,
including being used as stand-alone tools.
In particular, we have recently instrumented the SMT solver \cvc~\cite{CVC4-CAV-11} with
new capabilities which make it efficient for synthesis conjectures,
and entered it in the past several editions of the syntax-guided synthesis competition~\cite{alur2016results,DBLP:journals/corr/AlurFSS16a}
where it placed first in a number of categories.
This work has shown that
SMT solvers can be powerful tools
for handling synthesis conjectures using two orthogonal techniques:

\begin{enumerate}
\item {\bf Synthesis via Quantifier Instantiation}
The first leverages the support in SMT solvers
for first-order quantifier instantiation,
which has been used successfully in a number of approaches for automated theorem proving~\cite{ganzinger2003new,MouraBjoerner07EfficientEmatchingSmtSolvers,InstLA2016}.
Quantifier-instantiation techniques developed for SMT can be extended and used as a complete procedure for certain classes of synthesis conjectures.

\item {\bf Synthesis via Syntax-Guided Enumeration}
The second technique follows the syntax-guided synthesis paradigm~\cite{AlurETAL13SyntaxguidedSynthesis}.
More specifically, it simulates an enumerative search strategy
in the core of the SMT solver 
by leveraging its native support for algebraic datatypes.
\end{enumerate}

This paper gives an overview of the way these synthesis techniques can be embedded 
in the core of SMT solvers based on the DPLL(T) framework~\cite{nieuwenhuis2006solving}.
We focus on recent advances for achieving efficiency while giving 
consideration to the quality of solutions generated using both these techniques.

\paragraph{Overview}
In Section~\ref{sec:prelim},
we give preliminary definitions and introduce several 
classes of synthesis conjectures.
In Section~\ref{sec:synth}, we introduce
the concept of refutation-based synthesis and summarize
the scenarios in which it can be applied to synthesis conjectures.
In Section~\ref{sec:synth-qi},
we describe a synthesis technique based on quantifier instantiation,
its properties, and associated challenges.
In Section~\ref{sec:synth-sge},
we describe current work on developing syntax-guided synthesis techniques
in the core of an SMT solver.
In Section~\ref{sec:futurework},
we conclude with several directions for future work.

\section{Preliminaries}
\label{sec:prelim}

We consider synthesis in the context of 
a background theory $T$ in many-sorted logic.
Formally, a theory is a pair $( \Sigma, \mods )$
where $\Sigma$ is a signature and $\mods$ is a class of $\Sigma$-interpretations,
the intended models for $T$.
A formula is \define{$T$-satisfiable} (respectively, \define{$T$-valid}, \define{$T$-unsatisfiable})
if it is satisfied by some (respectively, every, no) interpretation in $\mods$.
A \define{synthesis conjecture} is a
formula of the form $\exists \vec{f}\, \forall \vec{x}\, P[ \vec{f}, \vec{x} ]$
with $\vec{f} = (f_1, \ldots, f_n)$ and $\vec x = (x_1, \ldots, x_k)$,
where
each $f_i$ in $\vec{f}$ has type of the form $\tau_1 \times \ldots \times \tau_{n_i} \rightarrow \tau$,
where $\tau_1,\ldots,\tau_{n_i},\tau$ are (first-order) sorts in $\Sigma$
and $P$ is a \emph{first-order} $\Sigma$-formula,
which means that in $P$ the second-order variables $\vec{f}$ are fully applied 
to arguments.
Here,
we write $P[ \vec{f}, \vec{x} ]$ to denote that the free variables
of formula $P$ are a subset of those in tuples $\vec{f}$ and $\vec{x}$,
and use this convention throughout the paper.
A \define{solution} to the synthesis conjecture is a substitution of the form
$\{f_1 \mapsto \lambda \vec{y}_1\, t_1,\, \ldots,\, f_n \mapsto \lambda \vec{y}_n\, t_n\}$
where $\vec{y}_1, \ldots, \vec{y}_n$ are bound variables and
$t_1, \ldots, t_n$ are $\Sigma$-terms with no second-order variables\footnote{%
In particular, it contains no variables from $\vec f$.
}
such that $\forall \vec{x}\, P[ (\lambda \vec{y}_1\, t_1, \ldots, \lambda \vec{y}_n\, t_n), \vec{x} ]$,
after $\beta$-reduction, is $T$-valid in $T$.
Note that we restrict our consideration to solutions were none of the $f_i$ need 
to be defined recursively.

Optionally, we may be interested
in synthesis conjectures in the presence of \define{syntactic restrictions}.
We specify syntactic restrictions by a grammar $\R = ( s_0, S, R )$ 
where $s_0$ is an initial symbol, $S$ is a set of symbols with $s_0 \in S$,
and $R$ is a set of rules of the form
$s \rightarrow t$, where $s \in S$ and $t$ is a term
built from the symbols in the signature of theory $T$, 
free variables, and symbols from $S$.
The rules define a rewrite relation over such terms, also denoted by $\rightarrow$, 
as expected.
We say a term $t$ is \define{generated} by $\R$ if 
$s_0 \rightarrow^\ast t$ where $\rightarrow^\ast$ is the reflexive-transitive closure 
of $\rightarrow$ and $t$ does not contain symbols from $S$.
For example, the terms $x$, $( x + x )$ and $( ( 1 + x ) + 1)$ are all generated by the grammar 
$\R = ( \gI, \{ \gI \}, \{ \gI \rightarrow x, \gI \rightarrow 1, \gI \rightarrow (\gI + \gI) \} )$.
We will write a grammar like this in BNF-style as:
\[
\begin{array}{l@{\quad}l@{\quad}l}
\gI & \rightarrow & x \spl 1 \spl (\gI + \gI)
\end{array}
\]
We say that a solution $\lambda \vec x\, t$ meets the syntactic restrictions
of $\R$ if $t$ is generated by it.

%In the remainder of the paper, we fix a theory $T$.

\subsection{Classes of Synthesis Conjectures}

We describe SMT approaches for synthesis
specialized to particular classes of conjectures.
We introduce the following terminology for defining those classes
with respect to a given $\Sigma$-theory $T$.

\begin{dfn}[Input-Output Example Conjectures]
\label{dfn:ioex-conj}
An input-output example synthesis conjecture is a formula of the form:
\begin{eqnarray*}
  \exists \vec{f}\,\forall \vec x\, ( \bigwedge_{k=0}^n \vec x \teq \vec i_k \Rightarrow \vec f( \vec x ) \teq \vec o_k )
\end{eqnarray*}
where for each $k=1,\ldots,n$, $\vec i_k$ and $\vec o_k$ are tuples of constants
from $\Sigma$.
\end{dfn}

\begin{example}\em
If $T$ is the theory of integer arithmetic with the usual signature, the formula:
\begin{eqnarray*}
  \exists f\,\forall x\, ( x \teq 1 \Rightarrow f( x ) \teq 2 ) \wedge 
                         ( x \teq 2 \Rightarrow f( x ) \teq 3 ) \wedge 
                         ( x \teq 7 \Rightarrow f( x ) \teq 8 )
\end{eqnarray*}
is an input-output example conjecture.
\qed
\end{example}

\begin{dfn}[Single invocation Conjectures]
A single invocation conjecture is a formula of the form:
\begin{eqnarray*}
  \exists \vec{f}\,\forall \vec x\, P[ \vec{f}( \vec x ), \vec x ]
\end{eqnarray*}
where $P[ \vec{f}( \vec x ), \vec x ]$ is an instance of a formula 
$P[ \vec y, \vec x ]$ that does not contain any second-order variables.
In other words, single invocation conjectures are those where all
functions in $\vec f$ are applied to the same argument tuple $\vec x$.
\end{dfn}

Notice that input-output example conjectures are a subset of single invocation conjectures.

\begin{example}\em
The formula:
\begin{eqnarray*}
  \exists f\,\forall x\,y\, ( f(x,y) \geq x \wedge f(x,y) \geq y )
\end{eqnarray*}
stating that $f$ returns a value greater than its arguments
is a single invocation synthesis conjecture.
\qed
\end{example}

All other synthesis conjectures that do not meet the criteria of the above definition 
we refer to as \emph{non-single invocation} conjectures.

\begin{example}\em
The formula:
\begin{eqnarray} \label{eqn:ioexampleex}
  \exists f\,\forall x\,y\, ( f(x,y) \teq f( y, x ) )
\end{eqnarray}
stating that $f$ is a commutative function 
is a non-single invocation synthesis conjecture.
\qed
\end{example}

\section{Refutation-Based Synthesis}
\label{sec:synth}

\begin{figure}[t]
  \centering
  \begin{tabular}{|c|c|c|c|}
  \hline
    \bf
    Syntax $\backslash$ Conjecture  & I/O Examples &  Single Invocation & Non Single Invocation  \\[0.5ex] \hline
    \multirow{2}{*}{restricted}   &    Enumerative + & Enumerative, or & \multirow{2}{*}{Enumerative} \\
                                    &  I/O sym breaking & CEGQI+reconstruction & \\[1ex]
    \hline
    \multirow{2}{*}{unrestricted}     &    CEGQI & \multirow{2}{*}{CEGQI} & Enumerative \\
                                    &  (trivially) & & (using default restrctions) \\
  \hline
  \end{tabular}
  \caption{Refutation-based techniques used by SMT solvers for solving classes of synthesis conjectures.}
    \label{fig:approaches}
\end{figure}

We use a \emph{refutation-based approach} for synthesis in SMT solvers~\cite{ReynoldsFMSD2017}
that takes as input the negation of a synthesis conjecture
$\neg \exists \vec{f}\, \forall \vec{x}\, P[ \vec{f}, \vec{x} ]$,
which is equivalent to 
$\forall \vec{f}\, \exists \vec{x}\, \neg P[ \vec{f}, \vec{x} ]$.
In this approach,
a solution for $\vec{f}$ can be extracted
from a proof of unsatisfiability\footnote{%
We will show examples of how solutions are extracted later in the paper.},
whereas a satisfiable response from the solver
indicates that the synthesis conjecture has no solutions.
This paper will focus solely on the former case,
that is, we consider only synthesis conjectures that have solutions.

Figure~\ref{fig:approaches} summarizes
the refutation-based techniques we use in DPLL(T)-based SMT solvers
for handling various classes of synthesis conjectures,
both with and without syntactic restrictions.
We give details on variants of counterexample-guided quantifier instantiation (CEGQI)
and syntax-guided enumerative search in the remainder of the paper.
As indicated in the figure,
counterexample-guided quantifier instantiation (Section~\ref{sec:synth-qi})
is applicable to single invocation conjectures only.
It also trivially applies to input-output examples without syntactic restrictions,
which we describe in Section~\ref{sec:cegqi-ioe}.
It typically is used only when no syntactic restrictions are associated with the conjecture,
although Section~\ref{sec:synth-qi-rcons} describes a technique
for reconstructing solutions from counterexample-guided instantiation that satisfy syntactic restrictions.
For other classes of conjectures, we use syntax-guided enumerative search (Section~\ref{sec:synth-sge}).
In the case where a synthesis conjecture is not single invocation but has 
no syntactic restrictions, we use a set of \emph{default restrictions}, 
that is, those that allow
all constructable $\Sigma$-terms as solutions.
For input-output examples, we may use a technique
for breaking symmetries in the search space based on evaluating input-output examples
which we describe in Section~\ref{sec:sb-ioee}.

\section{Synthesis via Counterexample-Guided Quantifier Instantiation}
\label{sec:synth-qi}

In previous work~\cite{ReynoldsDKBT15Cav}, we developed an
efficient technique for single invocation synthesis conjectures
without syntactic restrictions.
In this section, we give a brief review of this technique
and mention current challenges associated with this approach.
Unless otherwise stated, in all examples we will use linear integer arithmetic
as the background theory $T$ with sort $\sint$ for the set of all integers.
We will write $t_1 > t_2 > t_3$ as an abbreviation for $t_1 > t_2 \land t_2 > t_3$.

\begin{example}\em
\label{ex:qi-si}
Consider the synthesis conjecture:
\begin{eqnarray} \label{eqn:pre-si}
  \exists f\,\forall x\,y\, ( x > y+1 \Rightarrow x > f(x,y) > y ) \wedge ( y > x+1 \Rightarrow y > f(x,y) > x )
\end{eqnarray}
where $f$ is of type $\sint \times \sint \rightarrow \sint$ and all other variables are of type $\sint$.
This conjecture is single invocation and states that $f$ is a function that, 
under certain conditions on the inputs $x$ and $y$, returns a value
strictly between those inputs.
As noted by Reynolds et al.~\cite{ReynoldsDKBT15Cav}, the second-order formula above is equivalent
to the first-order formula:
\begin{eqnarray} \label{eqn:post-si}
  \forall x\,y\,\exists z\, ( x > y+1 \Rightarrow x > z > y ) \wedge ( y > x+1 \Rightarrow y > z > x )
\end{eqnarray}
where $z$ is of type $\sint$.
%This transformation is sometimes referred to as \emph{anti-skolemization}.
This formula states that for every two values $x$ and $y$ satisfying
certain restrictions, there exists a ``return'' value $z$ that is strictly 
between those values.

In contrast to formula (\ref{eqn:pre-si}), formula (\ref{eqn:post-si})
can be processed with SMT techniques for first-order quantified linear 
arithmetic~\cite{Bjoerner10LinearQuantifierEliminationAsAbstractDecision,DBLP:conf/lpar/BjornerJ15,dutertresolving,DBLP:conf/ijcai/FarzanK16,InstLA2016}.
In particular, since these techniques are refutation-based,
we consider the negation of (\ref{eqn:post-si}):
\begin{eqnarray} \label{eqn:post-si-neg}
  \exists xy\,\forall z\, \neg ( ( x > y+1 \Rightarrow x > z > y ) \wedge ( y > x+1 \Rightarrow y > z > x ) )
\end{eqnarray}
Using quantifier instantiation,
we can show that the instances of the innermost quantified formula:
\[\begin{array}{l}
  \neg ( ( x > y+1 \Rightarrow x > x+1 > y ) \wedge ( y > x+1 \Rightarrow y > x+1 > x ) ) \wedge {} \\
  \neg ( ( x > y+1 \Rightarrow x > y+1 > y ) \wedge ( y > x+1 \Rightarrow y > y+1 > x ) ) 
\end{array}\]
which simplify to $x > y+1$ and $y > x+1$ respectively,
are together $T$-unsatisfiable.
As described in~\cite{ReynoldsDKBT15Cav},
a solution for $f$ in (\ref{eqn:pre-si}) can be extracted
from the instantiations required for showing (\ref{eqn:post-si-neg}) $T$-unsatisfiable.
In particular, we construct a conditional function whose return values are
the terms we considered as instances of the negated first-order conjecture. 
In this case, from the above instances of (\ref{eqn:post-si-neg}) we construct 
the function
\begin{eqnarray*}
\lambda xy\, \ite( ( x > y+1 \Rightarrow x > x+1 > y ) \wedge ( y > x+1 \Rightarrow y > x+1 > x ),\, x+1,\, y+1 )
\end{eqnarray*}
which states that when the conjecture holds for $x+1$, return $x+1$,
otherwise return $y+1$.
After simplification, this gives the solution
$f = \lambda xy\, \ite( x \leq y+1,\, x+1,\, y+1 )$,
which indeed is a solution for our original conjecture in (\ref{eqn:pre-si}).
\qed
\end{example}

The key technical challenge in the previous example 
was to determine a $T$-unsatisfiable set of instances of (\ref{eqn:post-si-neg}).
We use a technique called \define{counterexample-guided quantifier instantiation} (CEGQI)
for determining these instances.
Variants of CEGQI have been used in 
a number of recent works~\cite{Monniaux10QuantifierEliminationLazyModelEnumeration,komuravelli2014smt,DBLP:conf/lpar/BjornerJ15,dutertresolving,DBLP:conf/lpar/FedyukovichGS15,InstLA2016}.
A detailed description of the technique can be found in~\cite{InstLA2016};
we summarize the most important details in the following.

For a negated single invocation synthesis conjecture
$\neg \exists \vec{f}\,\forall \vec x\, P[ \vec{f}( \vec x ), \vec x ]$,
our goal is to find a set $\Gamma$ of instances of the innermost body of the equivalent first-order formula
$\exists \vec x\, \forall \vec z\, \neg P[ \vec z , \vec x ]$ that are collectively $T$-unsatisfiable.
We incrementally construct the set
$\Gamma = \{ \neg P[ \vec t_1, \vec x ], \neg P[ \vec t_2, \vec x ],\ldots \}$
where each $\vec t_i$ is chosen by a \emph{selection function}.
%whose interface is given by the following definition.

\begin{dfn}[Selection Function]
A \emph{selection function} for $\forall \vec z\, \neg P[ \vec z , \vec x ]$
takes as input:
%\ct{check the corrections here}
\begin{enumerate}
\item a $\Sigma$-interpretation $\I$,
\item a tuple of fresh variables $\vec k$ with the same length and type as $\vec z$, and
\item a set of formulas 
$\Gamma = \{ \neg P[ \vec t_1, \vec x ], \ldots, \neg P[ \vec t_n, \vec x ], P[ \vec k, \vec x ] \}$ where $\I \models \Gamma$.
\end{enumerate}
It returns a tuple of terms $\vec t$ of the same type as $\vec k$ 
whose free variables are a subset of $\vec k$.
\end{dfn}

We use a selection function for finding the next instance $\neg P[ \vec t, \vec x ]$  
of $\neg P[ \vec z, \vec x ]$ to add to $\Gamma$.
Typically, the terms $\vec t$ are chosen based on the model $\I$ for $\Gamma$.
In particular, a model-based selection function is one that choses 
$\vec t$ based on the value of $\vec k$ in $\I$.
With sufficient conditions on the selection function used,
one may develop sound and complete instantiation procedures for the satisfiability
of $\exists\forall$-formulas in certain theories, which hence can be used
as sound and complete procedures for single invocation synthesis conjectures without syntactic restrictions.

Selection functions are specific to the background theory $T$
and are often inspired by quantifier elimination procedures.
Consider the case of a formula $\forall z\, \neg P[ z, \vec x ]$ with a single quantified variable (corresponding to an original synthesis conjecture involving a single function).
A selection function will chose instantiations for $z$ based on models $\I$ that satisfy $P[ k, \vec x ]$.
For linear real arithmetic,
a selection function for this formula may choose to return a tuple of terms
corresponding to the maximal lower bound (respectively minimal upper bound) for $k$ in $\I$,
where virtual terms such as $\infty$ and $\delta$ may be necessary.
This is analogous to the quantifier elimination procedure by Loos and Weispfenning~\cite{Loos93applyinglinear}.
We may also add the midpoint of the maximal lower and minimal upper bounds for $k$,
analogous to Ferrante and Rackoff's method~\cite{FerranteRackoff79ComputationalComplexityLogicalTheories}.
It is also possible to devise selection functions for linear integer arithmetic
that take into account implied divisibility constraints for $k$, similar to Cooper's method~\cite{cooper1972}.

%In practice, we develop selection functions based on quantifier elimination procedures.

We remark that there is a correspondence between three classes of procedures:
\begin{enumerate}
\item quantifier elimination procedures, 
\item instantiation-based procedures for $\exists \forall$ formulas, and
\item synthesis procedures for single invocation conjectures without syntactic restrictions.
\end{enumerate}
In particular, 
devising a sound and complete instantiation procedure for $\exists \forall$ formulas (Point 2)
is sufficient both for quantifier elimination (Point 1)
and for devising a sound and complete procedure for single invocation conjectures
without syntactic restrictions (Point 3)~\footnote{%
For details, see~\cite{InstLA2016}.
}.
Furthermore, a complete synthesis procedure for single invocation conjectures (Point 3)
is trivially sufficient for devising a complete instantiation procedure (Point 2).
Finally, many quantifier elimination techniques (Point 1), for instance based on virtual term substitution~\cite{Loos93applyinglinear},
can be rephrased as instantiation procedures (Point 2), although this direction of the correspondence
does not necessarily hold in general.
A number of previous works, e.g.~\cite{DBLP:conf/pldi/KuncakMPS10,sturm2011verification}, 
are based on the correspondence between quantifier elimination and synthesis.

\subsection{Inferring When a Conjecture is Single Invocation}

It is often the case that a conjecture is not single invocation
but is equivalent to one that is.
The latter can often be generated automatically from the former by a normalization
process that includes normalizing the arguments of invocations across conjunctions, and using quantifier elimination 
to eliminate variables for which
the function to synthesize is not applied.

\begin{example}\em
The (non-single invocation) synthesis conjecture $\exists f\, f( 0 ) \teq 1 \wedge f( 1 ) \teq 5$ is
equivalent to the single invocation conjecture $\exists f\, \forall x\, ( x \teq 0 \Rightarrow f( x ) \teq 1 ) \wedge ( x \teq 1 \Rightarrow f( x ) \teq 5 )$.
\qed
\end{example}

\begin{example}\em
Consider the conjecture:
\begin{eqnarray*} 
\exists f\, \forall x\,y\,z\, ( x \geq y \wedge x \teq z ) \vee ( y \geq x \wedge y \teq z ) ) \Rightarrow f( x, y ) \teq z
\end{eqnarray*}
where $f$ is of type $\sint \times \sint \rightarrow \sint$ and all other variables have type $\sint$.
Although $f$ is only invoked once, this conjecture does not fit 
the single invocation pattern due to the additional quantification on $z$.
However, quantification on $z$ may be eliminated based on the following steps.
First, replace the occurrence of $f( x, y )$ with a fresh variable $w$, and negate the conjecture to obtain:
\begin{eqnarray*} 
(\forall w\ \exists x\,y\,) \exists z\ \neg ( ( x \geq y \wedge x \teq z ) \vee ( y \geq x \wedge y \teq z ) ) \Rightarrow w \teq z )\ .
\end{eqnarray*}
As this is a formula in linear arithmetic, we may use a quantifier elimination procedure to obtain an 
equivalent formula involving only $w$, $x$, and $y$ such as:
\begin{eqnarray*} 
\forall w\ \exists x\,y\, \neg ( ( x \geq y \Rightarrow w \teq x ) \wedge ( y \geq x \Rightarrow w \teq y ) )\ .
\end{eqnarray*}
From that, we can generate the following single invocation conjecture,
which is provably equivalent to the original one:
\begin{eqnarray*} 
\exists f\, \forall x\,y\, ( x \geq y \Rightarrow f( x, y ) \teq x ) \wedge ( y \geq x \Rightarrow f( x, y ) \teq y ) \ .
\end{eqnarray*}
\qed
\end{example}

\subsection{Counterexample-Guided Quantifier Instantiation for I/O Examples}
\label{sec:cegqi-ioe}

While counterexample-guided quantifier instantiation
is both highly efficient and complete for single invocation synthesis conjectures,
it has the disadvantage of producing solutions that are suboptimal 
in terms of term size, especially in the case of partial specifications.

To see that, consider that based on Definition~\ref{dfn:ioex-conj}, all
input-output example synthesis conjectures
are also single invocation conjectures;
thus, the techniques above are applicable,
However, as demonstrated in the following example,
synthesis by counterexample-guided quantifier instantiation produces 
for this class of conjectures
sub-optimal solutions that, in a sense, overfit the specification.

\begin{example}\em
Consider the negated input-output example conjecture:
\begin{eqnarray} \label{eqn:ioe-ex}
  \neg \exists f\,\forall x\, ( x \teq 1 \Rightarrow f( x ) \teq 2 ) \wedge 
                         ( x \teq 2 \Rightarrow f( x ) \teq 3 ) \wedge 
                         ( x \teq 7 \Rightarrow f( x ) \teq 8 )\ .
\end{eqnarray}
This formula is equivalent to the first-order formula:
\begin{eqnarray*} 
  \exists x\,\forall z\,\neg( ( x \teq 1 \Rightarrow z \teq 2 ) \wedge 
                              ( x \teq 2 \Rightarrow z \teq 3 ) \wedge 
                              ( x \teq 7 \Rightarrow z \teq 8 ) )
\end{eqnarray*}
which can be shown $T$-unsatisfiable with the following three instances:
\[\begin{array}{l} 
  \neg( ( x \teq 1 \Rightarrow 2 \teq 2 ) \wedge 
        ( x \teq 2 \Rightarrow 2 \teq 3 ) \wedge 
        ( x \teq 7 \Rightarrow 2 \teq 8 ) ), \\
  \neg( ( x \teq 1 \Rightarrow 3 \teq 2 ) \wedge 
        ( x \teq 2 \Rightarrow 3 \teq 3 ) \wedge 
        ( x \teq 7 \Rightarrow 3 \teq 8 ) ), \\
  \neg( ( x \teq 1 \Rightarrow 8 \teq 2 ) \wedge 
        ( x \teq 2 \Rightarrow 8 \teq 3 ) \wedge 
        ( x \teq 7 \Rightarrow 8 \teq 8 ) )
\end{array}\]
which simplify to 
$x \teq 2 \vee x \teq 7$, $x \teq 1 \vee x \teq 7$ and $x \teq 1 \vee x \teq 2$, respectively.
Hence:
\begin{eqnarray*} 
\lambda x\, \ite( 
\begin{tabular}{c}
$( x \teq 1 \Rightarrow 2 \teq 2 ) \wedge$ \\
$( x \teq 2 \Rightarrow 2 \teq 3 ) \wedge$ \\
$( x \teq 7 \Rightarrow 2 \teq 8 ) \phantom{\wedge}$
\end{tabular}
,\; 2,\; \ite( 
\begin{tabular}{c}
$( x \teq 1 \Rightarrow 3 \teq 2 ) \wedge$ \\
$( x \teq 2 \Rightarrow 3 \teq 3 ) \wedge$ \\
$( x \teq 7 \Rightarrow 3 \teq 8 ) \phantom{\wedge}$  
\end{tabular}
,\; 3,\; 8 ) )
\end{eqnarray*}
which simplifies to $\lambda x\, \ite( x \teq 1,\, 2,\, \ite( x \teq 2,\, 3,\, 8 ) )$
is a solution for $f$ in (\ref{eqn:ioe-ex}).
In other words, counter\-example-guided quantifier instantiation
produces the trivial solution (consisting of an input-output table)
for the given input-output example conjecture
although shorter solutions, such as $\lambda x\, x+1$, exist.
\qed
\end{example}

From this example, one can see that a more compelling use case
of input-output example conjectures is when syntactic restrictions are imposed 
on the conjecture, which may force the synthesis procedure to produce more compact solutions.
Indeed, the programming-by-examples paradigm~\cite{DBLP:conf/popl/Gulwani11} assumes 
such restrictions are provided
so that intended solutions are generated by the underlying synthesis algorithm.
There, the additional goal is also to capture function intended by the user
by generalizing from a few input-output examples.

\subsection{Counterexample-Guided Quantifier Instantiation with Syntactic Restrictions}
\label{sec:synth-qi-rcons}

To find solutions for single invocation synthesis conjectures 
in the presence of syntactic restrictions,
a straightforward if incomplete technique is to first solve for the conjecture
as before, ignoring the restrictions, and then check 
whether the generated solution satisfies the syntactic restrictions.
If it does not, then we may try to reconstruct those portions of 
the solution that break the restrictions.  
We demonstrate this process in the following example.

\begin{example}\em
Consider again the synthesis conjecture from Example~\ref{ex:qi-si}
but now assume we are additionally given syntactic restrictions expressed by
this grammar with initial symbol $\gI$:
\[
\begin{array}{l@{\quad}l@{\quad}l}
\gI & \rightarrow & 0 \spl 1 \spl x \spl y \spl \gI + \gI \spl \ite( \gB, \gI, \gI ) \\
\gB & \rightarrow & \gI > \gI \spl \gI \teq \gI \spl \neg ( \gB )
\end{array}
\]
We may first ignore syntactic restrictions and
run counterexample-guided quantifier instantiation on the conjecture,
obtaining the solution $f = \lambda xy\, \ite( x \leq y+1, x+1, y+1 )$ as before.
This solution does not meet the syntactic restrictions above 
since the subterm $x \leq y+1$ cannot be generated from $\gB$.
However, the term $\neg( x > y+1 )$, which \emph{is} generated from $\gB$,
is equivalent to $x \leq y+1$ in $T$.
It follows that $f = \lambda xy\, \ite( \neg (x > y+1), x+1, y+1 )$
is also a solution for (\ref{eqn:pre-si}), which moreover satisfies 
the given syntactic restrictions.
\qed
\end{example}

We refer to this technique as counterexample-guided quantifier instantiation
with \emph{solution reconstruction}.\footnote{%
For more details on this technique, see Section 5.2 of~\cite{ReynoldsDKBT15Cav}.
}
Due to its heuristic nature, a synthesis procedure
based on syntax-guided enumeration often has a higher success rate 
than one based on this approach.
Thus, for single invocation properties with syntactic restrictions,
we may use a portfolio approach that first tries counterexample-guided instantiation
but aborts if solution reconstruction does not quickly succeed,
and then resorts to syntax-guided enumeration, which we describe in detail in Section~\ref{sec:synth-sge}.

\subsection{Synthesis via Quantifier Instantiation for Other Theories}

While instantiation-based procedures for quantified linear arithmetic are now fairly mature~\cite{DBLP:conf/lpar/BjornerJ15,dutertresolving,InstLA2016},
procedures for quantified constraints in other theories are still undergoing rapid development.
Notably, current methods for quantified bit-vectors construct
streams of instantiations based on candidate models~\cite{wintersteiger2013efficiently}, 
and use aggressive rewriting techniques
to increase the likelihood that the problem can be solved during preprocessing.
The limitation of current procedures is that
they are often unable to construct useful symbolic instantiations required for finding concise proofs of unsatisfiability.
To address this issue,
a recent approach by Preiner et al.~\cite{DBLP:conf/tacas/PreinerNB17} uses syntax-guided synthesis
to find relevant instantiations for quantified bit-vectors.
An independent approach by Rabe et al.~\cite{DBLP:conf/sat/RabeS16} uses new techniques to
construct symbolic Skolem functions for quantified Boolean formulas.

Furthermore, a number of SMT solvers have been extended with
theories outside the standard canon of traditional theories,
such as unbounded strings and regular expressions~\cite{LiangRTBD14},
finite sets~\cite{BanEtAl-IJCAR-16},
and floating point arithmetic~\cite{brain2014deciding}.
Devising counterexample-guided instantiation procedures for each of these theories remains an open challenge.

\section{Synthesis via Syntax-Guided Enumeration}
\label{sec:synth-sge}

For some cases, counterexample-guided quantifier instantiation
is either not applicable to the class of synthesis conjecture under consideration,
or suboptimal because of user-provided syntactic restrictions.
In these cases, we use enumerative syntax-guided techniques 
popularized by a number of recent synthesis tools, 
notably the enumerative solver by Udupa et al.~\cite{udupa2013transit,alurscaling}.
At a high level, these techniques
consider a stream of candidate solutions
sorted increasingly with respect to some total ordering,
typically some linearization of term size.
In this section, we focus on how these techniques can be
performing using existing components of a DPLL(T)-based SMT solver.
We will use the following running example.

\begin{example}\em
\label{ex:sge}
Consider the negated synthesis conjecture:
\begin{eqnarray} \label{eqn:nsi}
  \neg \exists f\,\forall x\,y\, ( f(x,y) \geq x \wedge f(x,y) \teq f(y,x) )
\end{eqnarray}
with syntactic restrictions on $f$ provided by the following grammar with initial symbol $\gI$:
\[
\begin{array}{l@{\quad}l@{\quad}l}
\gI & \rightarrow & 0 \spl x \spl y \spl \gI + 1 \spl \ite( \gB, \gI, \gI ) \\
\gB & \rightarrow & \gI \leq \gI \spl \gI \geq \gI \spl \gI \teq \gI \spl \neg ( \gB )
\end{array}
\]
Since SMT solvers do not natively have a notion of grammars,
we rephrase the syntactic restrictions as a (set of) \emph{algebraic datatypes},
for which a number of SMT solvers have dedicated decision procedures~\cite{DBLP:journals/entcs/BarrettST07,reynolds2015decision}.
Hence we construct the following set of (mutually recursive) algebraic datatypes:
\[
\begin{array}{l@{\quad}l@{\quad}l}
\dI & = & \con{0} \spl \con{x} \spl \con{y} \spl \con{plus}( \dI, \dI_1 ) \spl \con{if}( \dB, \dI, \dI ) \\
\dI_1 & = & \con{1} \\
\dB & = & \con{leq}( \dI, \dI ) \spl \con{eq}( \dI, \dI ) \spl \con{not}( \dB )
\end{array}
\]
Here, the right hand side of each datatype lists the set of possible constructors for the datatype,
where each constructor symbol corresponds to a symbol in the theory of arithmetic.
Notice that this construction involved flattening, so that each construct consists
of exactly one symbol applied to a set of arguments.
Thus, the rule $\gI \rightarrow \gI + 1$ required the auxiliary 
datatype $\dI_1$ with a single constructor $\con{1}$.
The construction also involved minimization, and hence the rule
$\gB \rightarrow \gI \geq \gI$ was discarded since it is redundant with respect to
$\gB \rightarrow \gI \leq \gI$.
For a datatype constant $c$, we call \emph{analog of $c$} 
the term it corresponds to in the theory it encodes.
For example, the analog of $\con{plus}( \con{x}, \con{1} )$ is the arithmetic term $x+1$.

As described by Reynolds et al.~\cite{ReynoldsDKBT15Cav},
by using the datatypes $\dI, \dI_1, \dB$,
we can construct a first-order version of the formula (\ref{eqn:nsi}) that
encodes the syntactic restrictions on solutions for $f$.
To do that, for $\tau$ in $( \dI, \dI_1, \dB )$, 
we first introduce an operator $\eval_\tau$
or type $\tau \times \sint \times \sint \to \tau'$
where $\tau'$ is respectively in $( \sint, \sint, \sbool )$.
Informally, $\eval_\tau$ is interpreted as a function that takes a datatype value $d$, 
an integer value $a$ and an integer value $b$, and evaluates the analog of $d$ 
under the assignment $\{x \mapsto a, y \mapsto b\}$.
For example, 
$\eval_\dI( \con{x}, 2, 3 ) = 2$;
$\eval_\dI( \con{y}, 2, 3 ) = 3$;
$\eval_\dI( \con{plus}( \con{x}, \con{1} ), 2, 3 ) =$
$\eval_\dI( \con{x}, 2, 3 ) + \eval_{\dI_1}( \con{1}, 2, 3 ) =$
$2 + 1 = 3$;
$\eval_\dI( \con{if}(\con{leq}(\con{x},\con{y}),\con{1},\con{0}), 2, 3 ) =$
$\ite(\eval_\dB(\con{leq}(\con{x},\con{y}), 2, 3), \eval_\dI(\con{1}, 2, 3), \eval_\dI(\con{0}, 2, 3)) = 1$.
Concretely, these semantics can be enforced in the SMT solver 
by implementing additional inference rules 
for unfolding and eliminating occurrences of the three $\eval_\tau$ recursively
over the datatypes $\dI$, $\dI_1$ and $\dB$.

Using these operators, we can construct a first-order formula corresponding to (\ref{eqn:nsi}):
\begin{eqnarray} \label{eqn:nsi-enc}
  \neg \exists d_f\,\forall x\,y\, ( \eval_\dI(d_f,x,y) \geq x \wedge \eval_\dI(d_f,x,y) \teq \eval_\dI(d_f,y,x) )
\end{eqnarray}
where, however, $d_f$ is a \emph{first-order} variable of datatype sort $\dI$
(corresponding to the second-order variable $f$ of type $\sint \times \sint \to \sint$). 
This formula is equivalent to:
\begin{eqnarray} \label{eqn:nsi-enc-neg}
  \forall d_f\,\exists xy\, \neg ( \eval_\dI(d_f,x,y) \geq x \wedge \eval_\dI(d_f,x,y) \teq \eval_\dI(d_f,y,x) )
\end{eqnarray}
which can be shown $T$-unsatisfiable by a single instantiation that maps $d_f$ to $\con{if}( \con{leq}( \con{y}, \con{x} ), \con{x}, \con{y} )$.
It is possible to prove that then $\lambda xy\, \ite( y \leq x, x, y )$ is a solution 
for $f$ in (\ref{eqn:nsi}).
\qed
\end{example}

Like the previous section, the key technical challenge in the above technique is
determining the datatype term $\con{if}( \con{leq}( \con{y}, \con{x} ), \con{x}, \con{y} )$
used to instantiate the universal quantifier in (\ref{eqn:nsi-enc-neg}).
At a high level, this is done brute-force, by considering a stream of
candidate terms based on some ordering.\footnote{%
This ordering is determined,
for instance, by counting the number of non-nullary constructors in the terms.
}
Such candidates can be generated by an extension of the decision procedure
for quantifier-free equational constraints over algebraic datatypes
which is implementend in some SMT solvers~\cite{DBLP:journals/entcs/BarrettST07,reynolds2015decision}.
Before giving more detail on the candidate generation procedure, 
we must introduce some more notation.

If $d$ is a term of some datatype type $\dD$ and $\con{C}$ is a constructor of $\dD$,
we write $\isc{d}{\con{C}}$ to denote a predicate that is satisfied
exactly when $d$ is interpreted as a datatype value whose top symbol is $\con{C}$.
These predicates are sometimes referred to as \emph{discriminators}.
We write $\ssel{d}{\tau}{n}$ to denote the term that is interpreted as 
the $n^{th}$ child of $d$ that has type $\tau$ if one exists, or is freely interpreted otherwise.
In Example~\ref{ex:sge},
if $d^\I = \con{plus}( \con{x}, \con{1} )$, then
$\ssel{d}{\dI}{1}^\I = \con{x}$ and
$\ssel{d}{\dI_1}{1}^\I = \con{1}$, 
whereas if $d^\I = \con{if}( \con{leq}( \con{x}, \con{y} ), \con{y}, \con{x} )$, then
$\ssel{d}{\dB}{1}^\I = \con{leq}( \con{x}, \con{y} )$,
$\ssel{d}{\dI}{2}^\I = \ssel{\ssel{d}{\dB}{2}}{\dI}{1}^\I =\con{x}$ and
$\ssel{d}{\dI}{1}^\I = \ssel{\ssel{d}{\dB}{1}}{\dI}{2}^\I =\con{y}$.
The functions $\con{sel}_{\tau,n}$ are often referred to as \emph{selectors},
where here we allow that selector symbols may access the subfields of multiple constructors\footnote{%
In most presentation of algebraic datatypes~\cite{DBLP:journals/entcs/BarrettST07,reynolds2015decision}, selectors
are associated with one constructor only.
Decision procedures for datatypes can be easily modified 
to support our version of selectors and discriminators. 
}.

Now, consider again the algebraic datatypes $\dI, \dI_1, \dB$ from Example~\ref{ex:sge},
and let $d$ be a fresh variable of type $\dI$.
A DPLL(T)-based SMT solver may be configured to enumerate a stream of values of datatype $\dI$ 
by finding models $\I$ for a evolving set of clauses $\Gamma_d$, initially empty.
On each iteration, assuming $\Gamma_d$ is satisfied by some interpretation $\I$,
we consider the value of $d^\I$ as an instantiation for $d_f$ in (\ref{eqn:nsi-enc-neg}).
The value of $d^\I$ either corresponds to a solution for $f$ in (\ref{eqn:nsi}),
in which case the resulting instantiation simplifies to false,
or it does not, in which case the resulting instantiation simplifies to true.
In the latter case, we ensure all subsequent models $\J$ for $d$
are such that $d^\J \neq d^\I$.
To do so, we add a clause of the form
$\neg \isc{t_1}{\con{C_1}} \vee \ldots \vee \neg \isc{t_n}{\con{C_n}}$
where each $t_i$ is a (possibly empty) chain of selectors applied to $d$.

For example, if we find that the instantiation of (\ref{eqn:nsi-enc-neg}) 
that maps $d_f$ to $d^\I = \con{plus}( \con{x}, \con{1} )$ is not a solution,
then we add the clause:
\begin{eqnarray}
\neg \isc{d}{\con{plus}} \vee \neg \isc{\ssel{d}{\dI}{1}}{\con{x}} \vee \neg \isc{\ssel{d}{\dI_1}{2}}{\con{1}} 
\end{eqnarray}
to $\Gamma_d$.
This clause records the fact that we want
subsequent models $\J$ of $\Gamma_d$ to be such that $d^\J \neq \con{plus}( \con{x}, \con{1} )$,
since at least one of the disjuncts in the above formula must hold.

\subsection{Symmetry Breaking Based on Theory-Specific Simplification}

The key optimization for making enumerative syntax-guided search scalable in practice
is to avoid considering multiple solutions that are identifiable based on a suitable equivalence relation.
For this purpose,
we leverage the fact that DPLL(T)-based SMT solvers
use simplification techniques for terms and formulas.
Such techniques simplify the development of decision procedures
and can be used to improve the performance of subsolvers 
for certain background theories~\cite{DBLP:conf/cav/ReynoldsWBBLT17}.
The same simplification techniques can be used in turn to recognize when a candidate solution
to equivalent to another, thereby allowing us to prune portions
of the enumerative search.

Specifically, state-of-the-art SMT solvers are capable of constructing \
from any term $t$
a \define{simplified form}\footnote{%
This is sometimes also called its
normal or rewritten form.} that we denote here as $\nmf{t}$.
This is a term that is equivalent to $t$ in the background theory $T$, 
but is simpler by some measure.
We do not need to require simplified forms to be unique.
Specifically, equivalent terms $s$ and $t$ may have different simplified forms
$\nmf s$ and $\nmf t$.
However, the more equivalent terms that have the same simplified form,
the more effective our pruning technique is.
We illustrate how theory-specific simplification methods
can be used to prune search space in the following example.

\begin{example}\em
\label{ex:sym-break}
Consider the algebraic datatypes:
\[
\begin{array}{l@{\quad}l@{\quad}l}
\dI & = & \con{0} \spl \con{1} \spl \con{x} \spl \con{y} \spl \con{plus}( \dI, \dI ) \spl \con{if}( \dB, \dI, \dI ) \\
\dB & = & \con{leq}( \dI, \dI ) \spl \con{eq}( \dI, \dI ) \spl \con{not}( \dB )
\end{array}
\]
Given an interpretation $\I$, consider a list of candidate solutions $d^\I$, their arithmetic analog,
and their corresponding simplified form obtained with simplifications specific to integer arithmetic:
\[
\begin{tabular}{|c|c|c|c|}
\hline
  \# & $d^\I$  & Analog &  Analog$\downarrow$  \\[0.5ex] \hline
  \hline
  1 & $\con{x}$ & $x$ & $x$ \\
  & $\ldots$ & & \\
  2 & $\con{plus}( \con{y}, \con{1} )$ & $y+1$ & $1+y$ \\
  3 & $\con{plus}( \con{x}, \con{0} )$ & $x+0$ & $x$ \\
  4 & $\con{plus}( \con{1}, \con{y} )$ & $1+y$ & $1+y$ \\
  & $\ldots$ & & \\
  5 & $\con{if}( \con{leq}( \con{0}, \con{1} ), \con{x}, \con{0} )$ & $\ite( 0 \leq 1, x, 0 )$ & $x$ \\
\hline
\end{tabular}
\]
In this table, we assume our simplification orders
monomial sums based on some ordering so that, for instance, $\nmf{(y+1)} = 1+y$.
Notice that the third, fourth and fifth terms listed in this enumeration
have analogs that are identical after simplification to previous terms in the enumeration.
For example, the simplified analog of $\con{plus}( \con{x}, \con{0} )$ is the same as $\con{x}$.
For this reason, we may add clauses to the SMT solver that exclude models $\I$ where $d$ is interpreted as $\con{plus}( \con{x}, \con{0} )$.
In particular, the clauses:
\begin{eqnarray}
\label{eqn:ex-sym-break-clause}
\neg \isc{d}{\con{plus}} \vee \neg \isc{\ssel{d}{\dI}{1}}{\con{x}} \vee \neg \isc{\ssel{d}{\dI}{2}}{\con{0}} \\
\neg \isc{d}{\con{plus}} \vee \neg \isc{\ssel{d}{\dI}{1}}{\con{1}} \vee \neg \isc{\ssel{d}{\dI}{2}}{\con{y}} \\
\label{eqn:ex-sym-break-clause-end}
\begin{tabular}{r}
$\neg \isc{d}{\con{if}} \vee \neg \isc{\ssel{d}{\dB}{1}}{\con{leq}} 
\vee \neg \isc{\ssel{\ssel{d}{\dB}{1}}{\dI}{1}}{\con{0}} \vee \neg \isc{\ssel{\ssel{d}{\dB}{1}}{\dI}{2}}{\con{1}} \vee$ \\
$\neg \isc{\ssel{d}{\dI}{1}}{\con{x}} \vee \neg \isc{\ssel{d}{\dI}{2}}{\con{0}} \phantom{\vee}$
\end{tabular}
\end{eqnarray}
exclude models where $d$ is interpreted as $\con{plus}( \con{x}, \con{0} )$, $\con{plus}( \con{1}, \con{y} )$,
and $\con{if}( \con{leq}( \con{0}, \con{1} ), \con{x}, \con{y} )$ respectively.
We refer to formulas (\ref{eqn:ex-sym-break-clause}-\ref{eqn:ex-sym-break-clause-end}) as
\emph{symmetry breaking clauses}.
Notice that symmetry breaking clauses do not rule out interesting candidate solutions since they
only have the effect of avoiding repeated solutions modulo equivalence in $T$.
\qed
\end{example}

We extend the quantifier-free decision procedure for algebraic datatypes
to maintain a database of entries of form described in the table in Example~\ref{ex:sym-break}.
Symmetry breaking clauses are added to the set $\Gamma_d$ (introduced earlier) in two ways.
First, whenever a new datatype term $t$ is generated, we eagerly generate 
a predetermined set of symmetry breaking clauses to restrict the value of $t$ in all models,
where this set is determined by statically analyzing the set of constructors and the symbols in the theory they correspond to.
Second, when considering an interpretation $\I$, if any datatype term $t^\I$
has a value that is not unique with respect to existing values in our database,
we add a symmetry breaking clause that implies that $t \tneq t^\I$,
thereby forcing the solver to find a new interpretation.

Notice that symmetry breaking clauses can be reused for multiple terms of the same type.
For example, our implementation may consider
a modified version of
the symmetry breaking clause (\ref{eqn:ex-sym-break-clause})
where $d$ is replaced by $\ssel{d}{\dI}{1}$:
\begin{eqnarray}
\neg \isc{\ssel{d}{\dI}{1}}{\con{plus}} \vee \neg \isc{\ssel{\ssel{d}{\dI}{1}}{\dI}{1}}{\con{x}} \vee \neg \isc{\ssel{\ssel{d}{\dI}{1}}{\dI}{2}}{\con{0}}
\end{eqnarray}
This clause can be added to $\Gamma_d$, thereby enabling the solver to avoid candidate models where the first
child of $d$ with type $\dI$ is interpreted as $\con{plus}( \con{x}, \con{0} )$.
This further enables the solver to avoid candidate solutions like $\con{plus}( \con{plus}( \con{x}, \con{0} ), \con{y} )$.
A similar observation is made in the approach by Alur et al.~\cite{alurscaling}.

Symmetry breaking clauses may be generalized in several ways
to consider, for instance, cases when
a term simplifies to one of its proper subterms,
or when simplification can be performed independently of certain subterms.

\begin{example}\em
Recall that the symmetry breaking clause (\ref{eqn:ex-sym-break-clause})
in Example~\ref{ex:sym-break} was justified by the fact that
$\nmf{(x+0)} = x$, which has the same simplified form as the analog of $\con{x}$.
Since $\nmf{(z+0)} = z$ for fresh variable $z$,
we may conclude that $\nmf{(t+0)} = t$ for all terms $t$.
Thus, we may generalize this symmetry breaking clause 
to exclude the disjunct that depends on $\con{x}$, obtaining the clause:
\begin{eqnarray}
\neg \isc{d}{\con{plus}} \vee \neg \isc{\ssel{d}{\dI}{2}}{\con{0}}
\end{eqnarray}
This clause rules out all models where $d$ is interpreted as a value of the form
$\con{plus}( d_0, \con{0})$ \emph{for any} $d_0$.
It can be shown that this clause does not rule out any interesting candidate solutions,
and hence can be used to avoid an (infinite) class of candidate solutions.
\qed
\end{example}

\begin{example}\em
The symmetry breaking clause (\ref{eqn:ex-sym-break-clause-end})
in Example~\ref{ex:sym-break} was justified by the fact that
$\nmf{\ite( 0 \leq 1, x, 0 )} = x$.
We can generalize this observation to terms of the form
$\ite( 0 \leq 1, x, t )$
by noting that $\nmf{\ite( 0 \leq 1, x, z )} = x$ for fresh variable $z$. 
Thus, the simplification in this example did not depend on the second occurrence of $0$.
Hence, we can generalize (\ref{eqn:ex-sym-break-clause-end}) to exclude
its corresponding disjunct $\neg \isc{\ssel{d}{\dI}{2}}{\con{0}}$, obtaining the clause:
\begin{eqnarray}
\label{eqn:gen-sb-ii}
\neg \isc{d}{\con{if}} \vee \neg \isc{\ssel{d}{\dB}{1}}{\con{leq}} \vee
\neg \isc{\ssel{\ssel{d}{\dB}{1}}{\dI}{1}}{\con{0}} \vee \neg \isc{\ssel{\ssel{d}{\dB}{1}}{\dI}{2}}{\con{1}} \vee
\neg \isc{\ssel{d}{\dI}{1}}{\con{x}}
\end{eqnarray}
This clause rules out all models where $d$ is interpreted as a value of the form $\con{if}( \con{leq}( \con{0}, \con{1} ), \con{x}, d_0 )$ for any $d_0$.
We can generalize this further by noting
$\nmf{\ite( 0 \leq 1, z_1, z_2 )} = z_1$ for fresh variables $z_1, z_2$.
Hence
we may drop the disjunct $\neg \isc{\ssel{d}{\dI}{1}}{\con{x}}$ from (\ref{eqn:gen-sb-ii}) as well.
\qed
\end{example}

Generalizing symmetry breaking clauses has a dramatic positive impact 
on the performance of syntax-guided enumerative search in our implementation.
Notice that the above generalization techniques must take 
syntactic restrictions into account.
Consider the algebraic datatypes:
\[
\begin{array}{l@{\quad}l@{\quad}l}
\dI & = & \con{x} \spl \con{plus}( \dI_1, \dI_2 ) \\
\dI_1 & = & \con{0} \spl \con{1} \spl \con{x} \spl \con{y} \spl \con{plus}( \dI, \dI ) \\
\dI_2 & = & \con{0}
\end{array}
\]
Although the terms $\con{x}$ and $\con{plus}( \con{x}, \con{0} )$ 
have identical analogs after simplification, the symmetry breaking clause for $d : \dI$ of the form
$\neg \isc{d}{\con{plus}} \vee \neg \isc{\ssel{d}{\dI_1}{1}}{\con{x}} \vee \neg \isc{\ssel{d}{\dI_2}{1}}{\con{0}}$
cannot be generalized to
$\neg \isc{d}{\con{plus}} \vee \neg \isc{\ssel{d}{\dI_2}{1}}{\con{0}}$.
Even though $\nmf{(z+0)} = z$ for fresh variable $z$,
this clause clearly rules out possible solutions for $d$ like $\con{plus}( \con{y}, \con{0} )$
that could not be constructed if such a symmetry breaking clause were used.
In practice, we annotate terms with the datatype they correspond to.
In this example, $z + 0$ is annotated with $\dI$ and $z$ is annotated with $\dI_1$.
Hence, we can infer that the simplification $\nmf{(z+0)} = z$ does not preserve syntactic restrictions
since the annotated types of $z+0$ and $z$ are not identical.

\subsection{Symmetry Breaking Based on I/O Example Evaluation}
\label{sec:sb-ioee}

Alur et al.~\cite{alurscaling} note that enumerative syntax-guided 
search need only consider candidate solutions
that are unique when evaluated on the set of input points in the conjecture.
Thus, we may use an even stronger criterion for 
recognizing when candidate solutions can be discarded.
We demonstrate how
this observation can be incorporated in our setting
in the following example.

\begin{example}\em
\label{ex:sym-break-io}
Consider the input-output example conjecture:
\begin{equation} \label{eqn:ioe-ex}
\begin{array}{r@{\,}l}
 \neg \exists f\,\forall x\,y\ & ( x \teq 1 \wedge y \teq 0 \Rightarrow f( x, y ) \teq 1 ) \wedge {} \\
                              & ( x \teq 2 \wedge y \teq 1 \Rightarrow f( x, y ) \teq 3 ) \wedge {} \\
                               & ( x \teq 7 \wedge y \teq 1 \Rightarrow f( x, y ) \teq 8 )
\end{array}
\end{equation}
where $f$ is of type $\sint \times \sint \rightarrow \sint$ and all other 
variables are of type $\sint$.
Note that this conjecture consists of three conjunctions which correspond to
the input points $( x, y ) = ( 1,1 )$, $(2,1)$, and $(7,1)$.
Assume we are given syntactic restrictions for this conjecture,
which we transform into the algebraic datatype:
\[
\begin{array}{l@{\quad}l@{\quad}l}
\dI & = & \con{0} \spl \con{1} \spl \con{x} \spl \con{y} \spl \con{plus}( \dI, \dI ) \spl \con{if}( \dB, \dI, \dI ) \\
\dB & = & \con{leq}( \dI, \dI ) \spl \con{eq}( \dI, \dI ) \spl \con{not}( \dB )
\end{array}
\]
For each candidate solution $d^\I$, we again compute its arithmetic analog, and
its analog after theory-specific simplification; but now also compute 
its evaluation on the three input points of the conjecture:
\[
\begin{tabular}{|c|c|c|c|c|}
\hline
  \# & $d^\I$  & Analog &  Analog$\downarrow$ & Eval on Examples \\[0.5ex] \hline
  \hline
  1 & $\con{1}$ & $1$ & $1$ & $1,1,1$ \\
  2 & $\con{x}$ & $x$ & $x$ & $1,2,7$ \\
  & $\ldots$ & & &  \\
  3 & $\con{plus}( \con{x}, \con{x} )$ & $x+x$ & $1+y$ & $2,4,14$ \\
  4 & $\con{plus}( \con{x}, \con{0} )$ & $x+0$ & $x$ & n/a \\
  & $\ldots$ & & & \\
  5 & $\con{if}( \con{leq}( \con{y}, \con{x} ), \con{x}, \con{y} )$ & $\ite( y \leq x, x, y )$ & $\ite( y \leq x, x, y )$ & $1,2,7$ \\
  6 & $\con{if}( \con{leq}( \con{1}, \con{y} ), \con{1}, \con{x} )$ & $\ite( 1 \leq y, 1, x )$ & $\ite( 1 \leq y, 1, x )$ & $1,1,1$ \\
\hline
\end{tabular}
\]
We are interested in avoiding solutions
that are either not unique after simplification,
or not unique when evaluated on input examples.
Consider the fifth term $\con{if}( \con{leq}( \con{y}, \con{x} ), \con{x}, \con{y} )$,
whose analog after simplification is $\ite( y \leq x, x, y )$,
which is unique with respect to the previous terms listed in the enumeration.
However, evaluating $\ite( y \leq x, x, y )$ on the points
$( x, y ) = ( 1,1 )$, $(2,1)$, and $(7,1)$ gives $1$, $2$, and $7$ respectively.
Thus, with respect to the conjecture (\ref{eqn:ioe-ex}),
the candidate solutions $\con{x}$ and $\con{if}( \con{leq}( \con{y}, \con{x} ), \con{x}, \con{y} )$
are identical, and hence we may discard the latter using a symmetry breaking clause:
\begin{eqnarray}
\label{eqn:ex-sym-break-clause-io}
\begin{tabular}{r}
$\neg \isc{d}{\con{if}} \vee \neg \isc{\ssel{d}{\dB}{1}}{\con{leq}} 
\vee \neg \isc{\ssel{\ssel{d}{\dB}{1}}{\dI}{1}}{\con{y}} \vee \neg \isc{\ssel{\ssel{d}{\dB}{1}}{\dI}{2}}{\con{x}} \vee {}$ \\
$\neg \isc{\ssel{d}{\dI}{1}}{\con{x}} \vee \neg \isc{\ssel{d}{\dI}{2}}{\con{y}} \phantom{\vee {}}$
\end{tabular}
\end{eqnarray}
The candidate solutions $\con{1}$ and $\con{if}( \con{leq}( \con{1}, \con{y} ), \con{1}, \con{x} )$
are identical for similar reasons.
Observe that for the term $\con{plus}( \con{x}, \con{0} )$,
we find that its analog after simplification is equivalent to $\con{x}$, and hence
we do not need to compute its evaluation on the input examples
since they are guaranteed to be identical.
\qed
\end{example}

We may use symmetry breaking clauses to eliminate candidate solutions
that are not unique when considering input examples.
We may use the same enhancements from the previous section for these clauses as well,
namely, they can be reapplied to any term of the proper type
and generalized using similar techniques.
For example, the clause (\ref{eqn:ex-sym-break-clause-io}) can be generalized to:
\begin{eqnarray}
\neg \isc{d}{\con{if}} \vee \neg \isc{\ssel{d}{\dB}{1}}{\con{leq}} \vee
\neg \isc{\ssel{\ssel{d}{\dB}{1}}{\dI}{1}}{\con{y}} \vee \neg \isc{\ssel{\ssel{d}{\dB}{1}}{\dI}{2}}{\con{x}} \vee
\neg \isc{\ssel{d}{\dI}{1}}{\con{x}}
\end{eqnarray}
by noting that evaluating the term $\ite( y \leq x, x, z )$ on the points
$( x, y ) = ( 1,1 )$, $(2,1)$, and $(7,1)$ gives $1,2,7$ respectively for fresh variable $z$.
Hence we may drop the disjunct $\neg \isc{\ssel{d}{\dI}{2}}{\con{y}}$ 
since this reasoning did not depend on the second occurrence of $\con{y}$
in the candidate solution $\con{if}( \con{leq}( \con{y}, \con{x} ), \con{x}, \con{y} )$.

\section{Conclusions and Future Work}
\label{sec:futurework}

We have presented techniques for integrating techniques for syntax-guided 
synthesis in the core of a DPLL(T)-based SMT solver.
All techniques described in the paper are implemented in the open-source SMT solver \cvc.
For future work,
we are investigating cases where synthesis by
quantifier instantiation and synthesis by syntax-guided enumeration
can be combined, as well as investigating
efficient quantifier instantiation techniques for new background theories.

%One possible direction is to exploit the fact that 
%many synthesis conjectures can be partitioned in portions that a single invocation and those that are not.

\bibliographystyle{eptcs}
\bibliography{main}
\end{document}

%% file: macros.tex
\newcommand{\cvc}{\textsc{cvc}{\small 4}\xspace}

\newcommand{\teq}{\approx}

\newcommand{\define}[1]{\textsl{#1}}

\newcommand{\eval}{\mathsf{eval}}

\newcommand{\I}{\mathcal{I}}
\newcommand{\J}{\mathcal{J}}
\newcommand{\R}{\mathcal{R}}

\newcommand{\mods}{\mathbf{I}}

\newcommand{\con}[1]{\mathsf{#1}}

\newcommand{\ite}{\con{ite}}

\newcommand{\nmf}[1]{#1{\downarrow}}

\newcommand{\tneq}{\not\teq}

\newcommand{\sint}{\mathsf{Int}}

\newcommand{\sbool}{\mathsf{Bool}}

\newcommand{\gI}{\con{I}}
\newcommand{\gB}{\con{B}}
\newcommand{\dI}{\con{I}}
\newcommand{\dB}{\con{B}}
\newcommand{\dD}{\con{D}}
\newcommand{\spl}{\ \mid \ }
\newcommand{\isc}[2]{\con{is}_{#2}( #1 )}
\newcommand{\ssel}[3]{\con{sel}_{#2,#3}( #1 )}

%% file: main.bbl
\begin{thebibliography}{10}
\providecommand{\bibitemdeclare}[2]{}
\providecommand{\surnamestart}{}
\providecommand{\surnameend}{}
\providecommand{\urlprefix}{Available at }
\providecommand{\url}[1]{\texttt{#1}}
\providecommand{\href}[2]{\texttt{#2}}
\providecommand{\urlalt}[2]{\href{#1}{#2}}
\providecommand{\doi}[1]{doi:\urlalt{http://dx.doi.org/#1}{#1}}
\providecommand{\bibinfo}[2]{#2}

\bibitemdeclare{inproceedings}{AlurETAL13SyntaxguidedSynthesis}
\bibitem{AlurETAL13SyntaxguidedSynthesis}
\bibinfo{author}{Rajeev \surnamestart Alur\surnameend},
  \bibinfo{author}{Rastislav \surnamestart Bod{\'{\i}}k\surnameend},
  \bibinfo{author}{Garvit \surnamestart Juniwal\surnameend},
  \bibinfo{author}{Milo M.~K. \surnamestart Martin\surnameend},
  \bibinfo{author}{Mukund \surnamestart Raghothaman\surnameend},
  \bibinfo{author}{Sanjit~A. \surnamestart Seshia\surnameend},
  \bibinfo{author}{Rishabh \surnamestart Singh\surnameend},
  \bibinfo{author}{Armando \surnamestart Solar{-}Lezama\surnameend},
  \bibinfo{author}{Emina \surnamestart Torlak\surnameend} \&
  \bibinfo{author}{Abhishek \surnamestart Udupa\surnameend}
  (\bibinfo{year}{2013}): \emph{\bibinfo{title}{Syntax-guided synthesis}}.
\newblock In: {\sl \bibinfo{booktitle}{FMCAD}}, \bibinfo{publisher}{{IEEE}},
  pp. \bibinfo{pages}{1--17}.

\bibitemdeclare{inproceedings}{alur2016results}
\bibitem{alur2016results}
\bibinfo{author}{Rajeev \surnamestart Alur\surnameend}, \bibinfo{author}{Dana
  \surnamestart Fisman\surnameend}, \bibinfo{author}{Rishabh \surnamestart
  Singh\surnameend} \& \bibinfo{author}{Armando \surnamestart
  Solar{-}Lezama\surnameend} (\bibinfo{year}{2015}):
  \emph{\bibinfo{title}{Results and Analysis of SyGuS-Comp'15}}.
\newblock In: {\sl \bibinfo{booktitle}{Proceedings Fourth Workshop on
  Synthesis, {SYNT} 2015, San Francisco, CA, USA, 18th July 2015.}}, pp.
  \bibinfo{pages}{3--26}, \doi{10.4204/EPTCS.202.3}.

\bibitemdeclare{inproceedings}{DBLP:journals/corr/AlurFSS16a}
\bibitem{DBLP:journals/corr/AlurFSS16a}
\bibinfo{author}{Rajeev \surnamestart Alur\surnameend}, \bibinfo{author}{Dana
  \surnamestart Fisman\surnameend}, \bibinfo{author}{Rishabh \surnamestart
  Singh\surnameend} \& \bibinfo{author}{Armando \surnamestart
  Solar{-}Lezama\surnameend} (\bibinfo{year}{2016}):
  \emph{\bibinfo{title}{SyGuS-Comp 2016: Results and Analysis}}.
\newblock In: {\sl \bibinfo{booktitle}{Proceedings Fifth Workshop on Synthesis,
  SYNT@CAV 2016, Toronto, Canada, July 17-18, 2016.}}, pp.
  \bibinfo{pages}{178--202}, \doi{10.4204/EPTCS.229.13}.

\bibitemdeclare{inproceedings}{alurscaling}
\bibitem{alurscaling}
\bibinfo{author}{Rajeev \surnamestart Alur\surnameend}, \bibinfo{author}{Arjun
  \surnamestart Radhakrishna\surnameend} \& \bibinfo{author}{Abhishek
  \surnamestart Udupa\surnameend} (\bibinfo{year}{2017}):
  \emph{\bibinfo{title}{Scaling Enumerative Program Synthesis via Divide and
  Conquer}}.
\newblock In: {\sl \bibinfo{booktitle}{Tools and Algorithms for the
  Construction and Analysis of Systems (TACAS)}}, pp.
  \bibinfo{pages}{319--336}, \doi{10.1007/978-3-319-21690-4_26}.

\bibitemdeclare{inproceedings}{BanEtAl-IJCAR-16}
\bibitem{BanEtAl-IJCAR-16}
\bibinfo{author}{Kshitij \surnamestart Bansal\surnameend},
  \bibinfo{author}{Andrew \surnamestart Reynolds\surnameend},
  \bibinfo{author}{Clark~W. \surnamestart Barrett\surnameend} \&
  \bibinfo{author}{Cesare \surnamestart Tinelli\surnameend}
  (\bibinfo{year}{2016}): \emph{\bibinfo{title}{A New Decision Procedure for
  Finite Sets and Cardinality Constraints in {SMT}}}.
\newblock In: {\sl \bibinfo{booktitle}{Automated Reasoning - 8th International
  Joint Conference, {IJCAR} 2016, Coimbra, Portugal, June 27 - July 2, 2016,
  Proceedings}}, pp. \bibinfo{pages}{82--98},
  \doi{10.1007/978-3-319-40229-1_7}.

\bibitemdeclare{inproceedings}{CVC4-CAV-11}
\bibitem{CVC4-CAV-11}
\bibinfo{author}{Clark \surnamestart Barrett\surnameend},
  \bibinfo{author}{Christopher~L. \surnamestart Conway\surnameend},
  \bibinfo{author}{Morgan \surnamestart Deters\surnameend},
  \bibinfo{author}{Liana \surnamestart Hadarean\surnameend},
  \bibinfo{author}{Dejan \surnamestart Jovanovic\surnameend},
  \bibinfo{author}{Tim \surnamestart King\surnameend}, \bibinfo{author}{Andrew
  \surnamestart Reynolds\surnameend} \& \bibinfo{author}{Cesare \surnamestart
  Tinelli\surnameend} (\bibinfo{year}{2011}): \emph{\bibinfo{title}{{CVC4}}}.
\newblock In: {\sl \bibinfo{booktitle}{Computer Aided Verification - 23rd
  International Conference, {CAV} 2011, Snowbird, UT, USA, July 14-20, 2011.
  Proceedings}}, pp. \bibinfo{pages}{171--177},
  \doi{10.1007/978-3-642-22110-1_14}.

\bibitemdeclare{article}{DBLP:journals/entcs/BarrettST07}
\bibitem{DBLP:journals/entcs/BarrettST07}
\bibinfo{author}{Clark \surnamestart Barrett\surnameend}, \bibinfo{author}{Igor
  \surnamestart Shikanian\surnameend} \& \bibinfo{author}{Cesare \surnamestart
  Tinelli\surnameend} (\bibinfo{year}{2007}): \emph{\bibinfo{title}{An Abstract
  Decision Procedure for Satisfiability in the Theory of Recursive Data
  Types}}.
\newblock {\sl \bibinfo{journal}{Electr. Notes Theor. Comput. Sci.}}
  \bibinfo{volume}{174}(\bibinfo{number}{8}), pp. \bibinfo{pages}{23--37},
  \doi{10.1016/j.entcs.2006.11.037}.

\bibitemdeclare{inproceedings}{Bjoerner10LinearQuantifierEliminationAsAbstractDecision}
\bibitem{Bjoerner10LinearQuantifierEliminationAsAbstractDecision}
\bibinfo{author}{Nikolaj \surnamestart Bj{\o}rner\surnameend}
  (\bibinfo{year}{2010}): \emph{\bibinfo{title}{Linear Quantifier Elimination
  as an Abstract Decision Procedure}}.
\newblock In \bibinfo{editor}{J{\"{u}}rgen \surnamestart Giesl\surnameend} \&
  \bibinfo{editor}{Reiner \surnamestart H{\"{a}}hnle\surnameend}, editors: {\sl
  \bibinfo{booktitle}{IJCAR}}, {\sl \bibinfo{series}{LNCS}}
  \bibinfo{volume}{6173}, \bibinfo{publisher}{Springer}, pp.
  \bibinfo{pages}{316--330}, \doi{10.1007/978-3-642-14203-1_27}.

\bibitemdeclare{inproceedings}{DBLP:conf/lpar/BjornerJ15}
\bibitem{DBLP:conf/lpar/BjornerJ15}
\bibinfo{author}{Nikolaj \surnamestart Bj{\o}rner\surnameend} \&
  \bibinfo{author}{Mikol{\'{a}}s \surnamestart Janota\surnameend}
  (\bibinfo{year}{2015}): \emph{\bibinfo{title}{Playing with Quantified
  Satisfaction}}.
\newblock In: {\sl \bibinfo{booktitle}{20th International Conferences on Logic
  for Programming, Artificial Intelligence and Reasoning - Short Presentations,
  {LPAR} 2015, Suva, Fiji, November 24-28, 2015.}}, pp.
  \bibinfo{pages}{15--27}.
\newblock \urlprefix\url{http://www.easychair.org/publications/paper/252316}.

\bibitemdeclare{article}{brain2014deciding}
\bibitem{brain2014deciding}
\bibinfo{author}{Martin \surnamestart Brain\surnameend}, \bibinfo{author}{Vijay
  \surnamestart D’Silva\surnameend}, \bibinfo{author}{Alberto \surnamestart
  Griggio\surnameend}, \bibinfo{author}{Leopold \surnamestart
  Haller\surnameend} \& \bibinfo{author}{Daniel \surnamestart
  Kroening\surnameend} (\bibinfo{year}{2014}): \emph{\bibinfo{title}{Deciding
  Floating-point Logic with Abstract Conflict Driven Clause Learning}}.
\newblock {\sl \bibinfo{journal}{Formal Methods in System Design}}
  \bibinfo{volume}{45}(\bibinfo{number}{2}), pp. \bibinfo{pages}{213--245},
  \doi{10.1007/s10703-013-0203-7}.

\bibitemdeclare{inproceedings}{cooper1972}
\bibitem{cooper1972}
\bibinfo{author}{D.~C. \surnamestart Cooper\surnameend} (\bibinfo{year}{1972}):
  \emph{\bibinfo{title}{Theorem Proving in Arithmetic without Multiplication}}.
\newblock In: {\sl \bibinfo{booktitle}{Machine Intelligence, pages 91–100}}.

\bibitemdeclare{inproceedings}{dutertresolving}
\bibitem{dutertresolving}
\bibinfo{author}{Bruno \surnamestart Dutertre\surnameend}
  (\bibinfo{year}{2015}): \emph{\bibinfo{title}{Solving Exists/Forall Problems
  With Yices}}.
\newblock In: {\sl \bibinfo{booktitle}{Workshop on Satisfiability Modulo
  Theories}}.

\bibitemdeclare{inproceedings}{DBLP:conf/ijcai/FarzanK16}
\bibitem{DBLP:conf/ijcai/FarzanK16}
\bibinfo{author}{Azadeh \surnamestart Farzan\surnameend} \&
  \bibinfo{author}{Zachary \surnamestart Kincaid\surnameend}
  (\bibinfo{year}{2016}): \emph{\bibinfo{title}{Linear Arithmetic
  Satisfiability via Strategy Improvement}}.
\newblock In: {\sl \bibinfo{booktitle}{Proceedings of the Twenty-Fifth
  International Joint Conference on Artificial Intelligence, {IJCAI} 2016, New
  York, NY, USA, 9-15 July 2016}}, pp. \bibinfo{pages}{735--743}.
\newblock \urlprefix\url{http://www.ijcai.org/Abstract/16/110}.

\bibitemdeclare{inproceedings}{DBLP:conf/lpar/FedyukovichGS15}
\bibitem{DBLP:conf/lpar/FedyukovichGS15}
\bibinfo{author}{Grigory \surnamestart Fedyukovich\surnameend},
  \bibinfo{author}{Arie \surnamestart Gurfinkel\surnameend} \&
  \bibinfo{author}{Natasha \surnamestart Sharygina\surnameend}
  (\bibinfo{year}{2015}): \emph{\bibinfo{title}{Automated Discovery of
  Simulation Between Programs}}.
\newblock In: {\sl \bibinfo{booktitle}{Logic for Programming, Artificial
  Intelligence, and Reasoning - 20th International Conference, {LPAR-20} 2015,
  Suva, Fiji, November 24-28, 2015, Proceedings}}, pp.
  \bibinfo{pages}{606--621}, \doi{10.1007/978-3-662-48899-7_42}.

\bibitemdeclare{book}{FerranteRackoff79ComputationalComplexityLogicalTheories}
\bibitem{FerranteRackoff79ComputationalComplexityLogicalTheories}
\bibinfo{author}{Jeanne \surnamestart Ferrante\surnameend} \&
  \bibinfo{author}{Charles~W. \surnamestart Rackoff\surnameend}
  (\bibinfo{year}{1979}): \emph{\bibinfo{title}{The Computational Complexity of
  Logical Theories}}.
\newblock {\sl \bibinfo{series}{Lecture Notes in Mathematics}}
  \bibinfo{volume}{718}, \bibinfo{publisher}{Springer},
  \doi{10.1007/BFb0062845}.

\bibitemdeclare{inproceedings}{ganzinger2003new}
\bibitem{ganzinger2003new}
\bibinfo{author}{Harald \surnamestart Ganzinger\surnameend} \&
  \bibinfo{author}{Konstantin \surnamestart Korovin\surnameend}
  (\bibinfo{year}{2003}): \emph{\bibinfo{title}{New directions in
  instantiation-based theorem proving}}.
\newblock In: {\sl \bibinfo{booktitle}{Logic in Computer Science, 2003.}},
  \bibinfo{organization}{IEEE}, \doi{10.1109/LICS.2003.1210045}.

\bibitemdeclare{inproceedings}{DBLP:conf/popl/Gulwani11}
\bibitem{DBLP:conf/popl/Gulwani11}
\bibinfo{author}{Sumit \surnamestart Gulwani\surnameend}
  (\bibinfo{year}{2011}): \emph{\bibinfo{title}{Automating string processing in
  spreadsheets using input-output examples}}.
\newblock In: {\sl \bibinfo{booktitle}{Proceedings of the 38th {ACM}
  {SIGPLAN-SIGACT} Symposium on Principles of Programming Languages, {POPL}
  2011, Austin, TX, USA, January 26-28, 2011}}, pp. \bibinfo{pages}{317--330},
  \doi{10.1145/1926385.1926423}.

\bibitemdeclare{inproceedings}{gulwani2011synthesis}
\bibitem{gulwani2011synthesis}
\bibinfo{author}{Sumit \surnamestart Gulwani\surnameend},
  \bibinfo{author}{Susmit \surnamestart Jha\surnameend},
  \bibinfo{author}{Ashish \surnamestart Tiwari\surnameend} \&
  \bibinfo{author}{Ramarathnam \surnamestart Venkatesan\surnameend}
  (\bibinfo{year}{2011}): \emph{\bibinfo{title}{Synthesis of loop-free
  programs}}.
\newblock In: {\sl \bibinfo{booktitle}{Proceedings of the 32nd {ACM} {SIGPLAN}
  Conference on Programming Language Design and Implementation, {PLDI} 2011,
  San Jose, CA, USA, June 4-8, 2011}}, pp. \bibinfo{pages}{62--73},
  \doi{10.1145/1993498.1993506}.

\bibitemdeclare{incollection}{komuravelli2014smt}
\bibitem{komuravelli2014smt}
\bibinfo{author}{Anvesh \surnamestart Komuravelli\surnameend},
  \bibinfo{author}{Arie \surnamestart Gurfinkel\surnameend} \&
  \bibinfo{author}{Sagar \surnamestart Chaki\surnameend}
  (\bibinfo{year}{2014}): \emph{\bibinfo{title}{{SMT}-based Model Checking for
  Recursive Programs}}.
\newblock In: {\sl \bibinfo{booktitle}{Computer Aided Verification}},
  \bibinfo{publisher}{Springer International Publishing},
  \doi{10.1007/978-3-319-08867-9_2}.

\bibitemdeclare{inproceedings}{DBLP:conf/pldi/KuncakMPS10}
\bibitem{DBLP:conf/pldi/KuncakMPS10}
\bibinfo{author}{Viktor \surnamestart Kuncak\surnameend},
  \bibinfo{author}{Mika{\"{e}}l \surnamestart Mayer\surnameend},
  \bibinfo{author}{Ruzica \surnamestart Piskac\surnameend} \&
  \bibinfo{author}{Philippe \surnamestart Suter\surnameend}
  (\bibinfo{year}{2010}): \emph{\bibinfo{title}{Complete functional
  synthesis}}.
\newblock In: {\sl \bibinfo{booktitle}{Proceedings of the 2010 {ACM} {SIGPLAN}
  Conference on Programming Language Design and Implementation, {PLDI} 2010,
  Toronto, Ontario, Canada, June 5-10, 2010}}, pp. \bibinfo{pages}{316--329},
  \doi{10.1145/1806596.1806632}.

\bibitemdeclare{inproceedings}{LiangRTBD14}
\bibitem{LiangRTBD14}
\bibinfo{author}{Tianyi \surnamestart Liang\surnameend},
  \bibinfo{author}{Andrew \surnamestart Reynolds\surnameend},
  \bibinfo{author}{Cesare \surnamestart Tinelli\surnameend},
  \bibinfo{author}{Clark \surnamestart Barrett\surnameend} \&
  \bibinfo{author}{Morgan \surnamestart Deters\surnameend}
  (\bibinfo{year}{2014}): \emph{\bibinfo{title}{A {DPLL(T)} Theory Solver for a
  Theory of Strings and Regular Expressions}}.
\newblock In: {\sl \bibinfo{booktitle}{Computer Aided Verification - 26th
  International Conference, {CAV} 2014}}, pp. \bibinfo{pages}{646--662},
  \doi{10.1007/978-3-319-08867-9_43}.

\bibitemdeclare{article}{Loos93applyinglinear}
\bibitem{Loos93applyinglinear}
\bibinfo{author}{R{\"{u}}diger \surnamestart Loos\surnameend} \&
  \bibinfo{author}{Volker \surnamestart Weispfenning\surnameend}
  (\bibinfo{year}{1993}): \emph{\bibinfo{title}{Applying Linear Quantifier
  Elimination}}.
\newblock {\sl \bibinfo{journal}{Comput. J.}}
  \bibinfo{volume}{36}(\bibinfo{number}{5}), pp. \bibinfo{pages}{450--462},
  \doi{10.1093/comjnl/36.5.450}.

\bibitemdeclare{inproceedings}{Monniaux10QuantifierEliminationLazyModelEnumeration}
\bibitem{Monniaux10QuantifierEliminationLazyModelEnumeration}
\bibinfo{author}{David \surnamestart Monniaux\surnameend}
  (\bibinfo{year}{2010}): \emph{\bibinfo{title}{Quantifier Elimination by Lazy
  Model Enumeration}}.
\newblock In: {\sl \bibinfo{booktitle}{Computer Aided Verification, 22nd
  International Conference, {CAV} 2010, Edinburgh, UK, July 15-19, 2010.
  Proceedings}}, pp. \bibinfo{pages}{585--599},
  \doi{10.1007/978-3-642-14295-6_51}.

\bibitemdeclare{inproceedings}{MouraBjoerner07EfficientEmatchingSmtSolvers}
\bibitem{MouraBjoerner07EfficientEmatchingSmtSolvers}
\bibinfo{author}{Leonardo~Mendon{\c{c}}a \surnamestart de~Moura\surnameend} \&
  \bibinfo{author}{Nikolaj \surnamestart Bj{\o}rner\surnameend}
  (\bibinfo{year}{2007}): \emph{\bibinfo{title}{Efficient E-Matching for {SMT}
  Solvers}}.
\newblock In \bibinfo{editor}{Frank \surnamestart Pfenning\surnameend}, editor:
  {\sl \bibinfo{booktitle}{CADE}}, {\sl \bibinfo{series}{LNCS}}
  \bibinfo{volume}{4603}, \bibinfo{publisher}{Springer}, pp.
  \bibinfo{pages}{183--198}, \doi{10.1007/978-3-540-73595-3_13}.

\bibitemdeclare{article}{nieuwenhuis2006solving}
\bibitem{nieuwenhuis2006solving}
\bibinfo{author}{Robert \surnamestart Nieuwenhuis\surnameend},
  \bibinfo{author}{Albert \surnamestart Oliveras\surnameend} \&
  \bibinfo{author}{Cesare \surnamestart Tinelli\surnameend}
  (\bibinfo{year}{2006}): \emph{\bibinfo{title}{Solving {SAT} and {SAT} Modulo
  Theories: From an abstract Davis--Putnam--Logemann--Loveland procedure to
  DPLL(\emph{T})}}.
\newblock {\sl \bibinfo{journal}{J. {ACM}}}
  \bibinfo{volume}{53}(\bibinfo{number}{6}), pp. \bibinfo{pages}{937--977},
  \doi{10.1145/1217856.1217859}.

\bibitemdeclare{inproceedings}{DBLP:conf/tacas/PreinerNB17}
\bibitem{DBLP:conf/tacas/PreinerNB17}
\bibinfo{author}{Mathias \surnamestart Preiner\surnameend},
  \bibinfo{author}{Aina \surnamestart Niemetz\surnameend} \&
  \bibinfo{author}{Armin \surnamestart Biere\surnameend}
  (\bibinfo{year}{2017}): \emph{\bibinfo{title}{Counterexample-Guided Model
  Synthesis}}.
\newblock In: {\sl \bibinfo{booktitle}{Tools and Algorithms for the
  Construction and Analysis of Systems (TACAS)}}, pp.
  \bibinfo{pages}{264--280}, \doi{10.1007/978-3-662-48899-7_42}.

\bibitemdeclare{inproceedings}{DBLP:conf/sat/RabeS16}
\bibitem{DBLP:conf/sat/RabeS16}
\bibinfo{author}{Markus~N. \surnamestart Rabe\surnameend} \&
  \bibinfo{author}{Sanjit~A. \surnamestart Seshia\surnameend}
  (\bibinfo{year}{2016}): \emph{\bibinfo{title}{Incremental Determinization}}.
\newblock In: {\sl \bibinfo{booktitle}{Theory and Applications of
  Satisfiability Testing - {SAT} 2016 - 19th International Conference,
  Bordeaux, France, July 5-8, 2016, Proceedings}}, pp.
  \bibinfo{pages}{375--392}, \doi{10.1007/978-3-319-40970-2_23}.

\bibitemdeclare{inproceedings}{reynolds2015decision}
\bibitem{reynolds2015decision}
\bibinfo{author}{Andrew \surnamestart Reynolds\surnameend} \&
  \bibinfo{author}{Jasmin~Christian \surnamestart Blanchette\surnameend}
  (\bibinfo{year}{2016}): \emph{\bibinfo{title}{A Decision Procedure for
  (Co)datatypes in {SMT} Solvers}}.
\newblock In: {\sl \bibinfo{booktitle}{Proceedings of the Twenty-Fifth
  International Joint Conference on Artificial Intelligence, {IJCAI} 2016, New
  York, NY, USA, 9-15 July 2016}}, pp. \bibinfo{pages}{4205--4209}.
\newblock \urlprefix\url{http://www.ijcai.org/Abstract/16/631}.

\bibitemdeclare{inproceedings}{ReynoldsDKBT15Cav}
\bibitem{ReynoldsDKBT15Cav}
\bibinfo{author}{Andrew \surnamestart Reynolds\surnameend},
  \bibinfo{author}{Morgan \surnamestart Deters\surnameend},
  \bibinfo{author}{Viktor \surnamestart Kuncak\surnameend},
  \bibinfo{author}{Cesare \surnamestart Tinelli\surnameend} \&
  \bibinfo{author}{Clark~W. \surnamestart Barrett\surnameend}
  (\bibinfo{year}{2015}): \emph{\bibinfo{title}{Counterexample-Guided
  Quantifier Instantiation for Synthesis in {SMT}}}.
\newblock In: {\sl \bibinfo{booktitle}{Computer Aided Verification - 27th
  International Conference, {CAV} 2015, San Francisco, CA, USA, July 18-24,
  2015, Proceedings, Part {II}}}, pp. \bibinfo{pages}{198--216},
  \doi{10.1007/978-3-319-21668-3_12}.

\bibitemdeclare{article}{InstLA2016}
\bibitem{InstLA2016}
\bibinfo{author}{Andrew \surnamestart Reynolds\surnameend},
  \bibinfo{author}{Tim \surnamestart King\surnameend} \&
  \bibinfo{author}{Viktor \surnamestart Kuncak\surnameend}
  (\bibinfo{year}{2017}): \emph{\bibinfo{title}{Solving quantified linear
  arithmetic by counterexample-guided instantiation}}.
\newblock {\sl \bibinfo{journal}{Formal Methods in System Design}},
  \doi{10.1007/s10703-017-0290-y}.

\bibitemdeclare{article}{ReynoldsFMSD2017}
\bibitem{ReynoldsFMSD2017}
\bibinfo{author}{Andrew \surnamestart Reynolds\surnameend},
  \bibinfo{author}{Viktor \surnamestart Kuncak\surnameend},
  \bibinfo{author}{Cesare \surnamestart Tinelli\surnameend},
  \bibinfo{author}{Clark \surnamestart Barrett\surnameend} \&
  \bibinfo{author}{Morgan \surnamestart Deters\surnameend}
  (\bibinfo{year}{2017}): \emph{\bibinfo{title}{Refutation-based synthesis in
  SMT}}.
\newblock {\sl \bibinfo{journal}{Formal Methods in System Design}},
  \doi{10.1007/978-3-540-30142-4_22}.

\bibitemdeclare{inproceedings}{DBLP:conf/cav/ReynoldsWBBLT17}
\bibitem{DBLP:conf/cav/ReynoldsWBBLT17}
\bibinfo{author}{Andrew \surnamestart Reynolds\surnameend},
  \bibinfo{author}{Maverick \surnamestart Woo\surnameend},
  \bibinfo{author}{Clark \surnamestart Barrett\surnameend},
  \bibinfo{author}{David \surnamestart Brumley\surnameend},
  \bibinfo{author}{Tianyi \surnamestart Liang\surnameend} \&
  \bibinfo{author}{Cesare \surnamestart Tinelli\surnameend}
  (\bibinfo{year}{2017}): \emph{\bibinfo{title}{Scaling Up {DPLL(T)} String
  Solvers Using Context-Dependent Simplification}}.
\newblock In: {\sl \bibinfo{booktitle}{Computer Aided Verification - 29th
  International Conference, {CAV} 2017, Heidelberg, Germany, July 24-28, 2017,
  Proceedings, Part {II}}}, pp. \bibinfo{pages}{453--474},
  \doi{10.1007/978-3-319-63390-9_24}.

\bibitemdeclare{article}{SolarLezama13ProgramSketching}
\bibitem{SolarLezama13ProgramSketching}
\bibinfo{author}{Armando \surnamestart Solar{-}Lezama\surnameend}
  (\bibinfo{year}{2013}): \emph{\bibinfo{title}{Program sketching}}.
\newblock {\sl \bibinfo{journal}{{STTT}}}
  \bibinfo{volume}{15}(\bibinfo{number}{5-6}), pp. \bibinfo{pages}{475--495},
  \doi{10.1007/s10009-012-0249-7}.

\bibitemdeclare{inproceedings}{sturm2011verification}
\bibitem{sturm2011verification}
\bibinfo{author}{Thomas \surnamestart Sturm\surnameend} \&
  \bibinfo{author}{Ashish \surnamestart Tiwari\surnameend}
  (\bibinfo{year}{2011}): \emph{\bibinfo{title}{Verification and synthesis
  using real quantifier elimination}}.
\newblock In: {\sl \bibinfo{booktitle}{Symbolic and Algebraic Computation,
  International Symposium, {ISSAC} 2011 (co-located with {FCRC} 2011), San
  Jose, CA, USA, June 7-11, 2011, Proceedings}}, pp. \bibinfo{pages}{329--336},
  \doi{10.1145/1993886.1993935}.

\bibitemdeclare{inproceedings}{tiwari2015program}
\bibitem{tiwari2015program}
\bibinfo{author}{Ashish \surnamestart Tiwari\surnameend},
  \bibinfo{author}{Adri{\`{a}} \surnamestart Gasc{\'{o}}n\surnameend} \&
  \bibinfo{author}{Bruno \surnamestart Dutertre\surnameend}
  (\bibinfo{year}{2015}): \emph{\bibinfo{title}{Program Synthesis Using Dual
  Interpretation}}.
\newblock In: {\sl \bibinfo{booktitle}{Automated Deduction - {CADE-25} - 25th
  International Conference on Automated Deduction, Berlin, Germany, August 1-7,
  2015, Proceedings}}, pp. \bibinfo{pages}{482--497},
  \doi{10.1007/978-3-319-21401-6_33}.

\bibitemdeclare{inproceedings}{udupa2013transit}
\bibitem{udupa2013transit}
\bibinfo{author}{Abhishek \surnamestart Udupa\surnameend},
  \bibinfo{author}{Arun \surnamestart Raghavan\surnameend},
  \bibinfo{author}{Jyotirmoy~V. \surnamestart Deshmukh\surnameend},
  \bibinfo{author}{Sela \surnamestart Mador{-}Haim\surnameend},
  \bibinfo{author}{Milo M.~K. \surnamestart Martin\surnameend} \&
  \bibinfo{author}{Rajeev \surnamestart Alur\surnameend}
  (\bibinfo{year}{2013}): \emph{\bibinfo{title}{{TRANSIT:} specifying protocols
  with concolic snippets}}.
\newblock In: {\sl \bibinfo{booktitle}{{ACM} {SIGPLAN} Conference on
  Programming Language Design and Implementation, {PLDI} '13, Seattle, WA, USA,
  June 16-19, 2013}}, pp. \bibinfo{pages}{287--296},
  \doi{10.1145/2462156.2462174}.

\bibitemdeclare{inproceedings}{DBLP:conf/aips/WangDCK16}
\bibitem{DBLP:conf/aips/WangDCK16}
\bibinfo{author}{Yue \surnamestart Wang\surnameend}, \bibinfo{author}{Neil~T.
  \surnamestart Dantam\surnameend}, \bibinfo{author}{Swarat \surnamestart
  Chaudhuri\surnameend} \& \bibinfo{author}{Lydia~E. \surnamestart
  Kavraki\surnameend} (\bibinfo{year}{2016}): \emph{\bibinfo{title}{Task and
  Motion Policy Synthesis as Liveness Games}}.
\newblock In: {\sl \bibinfo{booktitle}{Proceedings of the Twenty-Sixth
  International Conference on Automated Planning and Scheduling, {ICAPS} 2016,
  London, UK, June 12-17, 2016.}}, p. \bibinfo{pages}{536}.

\bibitemdeclare{article}{wintersteiger2013efficiently}
\bibitem{wintersteiger2013efficiently}
\bibinfo{author}{Christoph~M \surnamestart Wintersteiger\surnameend},
  \bibinfo{author}{Youssef \surnamestart Hamadi\surnameend} \&
  \bibinfo{author}{Leonardo \surnamestart De~Moura\surnameend}
  (\bibinfo{year}{2013}): \emph{\bibinfo{title}{Efficiently Solving Quantified
  Bit-vector Formulas}}.
\newblock {\sl \bibinfo{journal}{Formal Methods in System Design}}
  \bibinfo{volume}{42}(\bibinfo{number}{1}), pp. \bibinfo{pages}{3--23},
  \doi{10.1007/s10703-012-0156-2}.

\end{thebibliography}
